\documentclass[twocolumn,showpacs,preprintnumbers,amsmath,amssymb,superscriptaddress,prl]{revtex4-1}
\usepackage{graphicx}
\usepackage{dcolumn}
\usepackage{bm}

\begin{document}

\title{Appearance of Flat Bands and Edge States in Boron-Carbon-Nitride Nanoribbons}

\author{Tomoaki Kaneko}
\thanks{current address:}
\affiliation{Computational Materials Science Unit, NIMS, 
Sengen 1-2-1, Tsukuba 305-0047, Japan}

\author{Kikuo Harigaya}
 \email{k.harigaya@aist.go.jp}
 \affiliation{Nanosystem Research Institute, AIST,
Higashi 1-1-1, Tsukuba, 305-8565, Japan}

\author{Hiroshi Imamura}
  \affiliation{Spintronics Research Center, AIST,
Umezono 1-1-1, Tsukuba 305-8568, Japan}

\date{\today}

\begin{abstract}
 Presence of flat bands and edge states at the Fermi level in graphene nanoribbons with zigzag edges is one of the most interesting and attracting properties of nanocarbon materials
 but it is believed that they are quite fragile states and disappear when B and N atoms are doped at around the edges.
In this paper,  we theoretically investigate electronic and magnetic properties of boron-carbon-nitride (BCN) nanoribbons with zigzag edges where the outermost C atoms on the edges are alternately replaced with B and N atoms using the first principles calculations.
We show that BCN nanoribbons have the flat bands and edge states at the Fermi level in both H$_2$ rich and poor environments.
The flat bands are similar to those at graphene nanoribbons with zigzag edges, but the distributions of charge and spin densities are different between them.
A tight binding model and the Hubbard model analysis show that the difference in the distribution of charge and spin densities is caused by the different site energies of B and N atoms compared with C atoms.
\end{abstract}


\maketitle

Graphene nanoribbons (GNRs) are nanometer width graphene (Gr) sheets \cite{Fujita1996jpsj,Nakada1996prb}, and fabricated by cutting Gr using electron-beam lithography \cite{Han2007prl}, unzipping carbon nanotubes \cite{Jiao2009nature,Kosynkin2009nature} and synthesized using bottom up process \cite{Cai2010nature}.
The electronic properties of GNRs strongly depend on the edge structures.
At the zigzag edges, presence of the so-called edge states and flat bands at the Fermi level were predicted \cite{Fujita1996jpsj,Nakada1996prb}. 
Recently, the presence of edge states in unzipped GNRs were experimentally confirmed by Tao {\it et al.} \cite{Tao2011nphys}.

In the honeycomb lattice, there are two nonequivalent sites in the unit cell, A and B sublattices, and
such sublattice structure plays decisive role for the electronic and magnetic properties of the flat bands and edge states.
In the tight binding model description, the electronic density are accumulated on the one sublattice including outermost C site but identically zero on the other sublattice \cite{Fujita1996jpsj}.
In the presence of the spin degree of freedom, ferrimagnetic ordering at each site was predicted using the Hubbard model calculation and the first-principles calculations \cite{Fujita1996jpsj,Okada2001prl,Son2006prl}.
Therefore, the edge states reflect the sublattice structure in the spacial distribution of electronic and spin densities.

On the other hand, boron-carbon-nitride (BCN) nanoribbons should show tunable electronic and magnetic properties by B and N doping, because B and N atoms act as acceptors and donors in Gr, respectively.
The electronic and magnetic properties of BCN nanoribbon in which C atoms are partially substituted with BN has been investigated by several authors \cite{Nakamura2005prb,Kan2008jcp,He2010apl,Basheer2011njp,Kan2008jcp,Lu2010apl,Kaneko2012submitted}.
The most of the work handled the hexagonal boron-nitride (BN) and Gr hybridized structures and half metallic or ferrimagnetic properties along the edges were reported by the several authors
\cite{Nakamura2005prb,Kan2008jcp,He2010apl,Basheer2011njp,Kan2008jcp,Lu2010apl}.
However, they could not obtain the flat bands and edge states in BCN nanoribbons where B and N atoms are doped near the edges because such doping disturbs the electronic structures.

\begin{figure*}[t!]
\centering
\includegraphics[width=15cm]{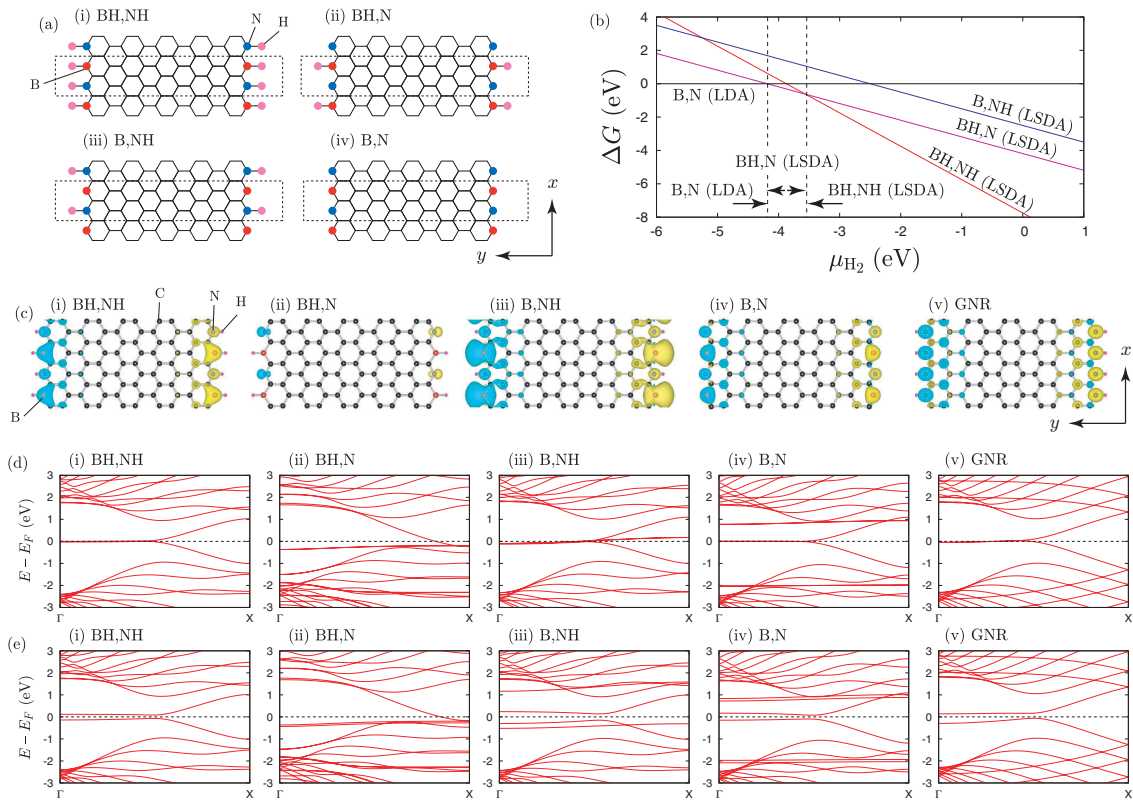}
\caption{(color online)
(a) Schematic illustration of the structure of zigzag BCN nanoribbons.
In this figure, the red, blue and pink circles indicate B, N and H atoms, respectively.
(b) The Gibbs free energy measured from that of B,N structure.
(c) The isosurface plots of spin distribution of BCN nanoribbons.
The isovaules are chosen as $\pm 1.0\times10^{-3}/{\rm bohr}^3$.
The black circles represent C atoms.
(d) The band structures of nonmagnetic BCN nanoribbons within LDA.
(e) The band structures of magnetized BCN nanoribbons within LSDA.
}
\label{fg:BN-alt-Fig1}
\end{figure*}

In this letter, we investigate the electronic and magnetic properties of BCN nanoribbons with zigzag edges where the outermost C atoms on the edge are alternately replaced with B and N atoms as shown in Fig.\ \ref{fg:BN-alt-Fig1} (a) using the first principles calculations.
We show that BCN nanoribbons in both H$_2$ rich or poor environment have the partially flat bands at the Fermi level.
We discuss the properties of edge states using the tight binding model and the Hubbard model calculations.

We performed the first-principles calculations based on the density functional theories within projector augmented wave method \cite{Blochl1994} and the local (spin) density approximation [L(S)DA] \cite{Perdew1981} implemented on {\scriptsize VASP} code \cite{Kresse1996prb,Kresse1996cms}.
We considered four different structures as shown in Fig.\ \ref{fg:BN-alt-Fig1} (a) in order to investigate the stability of H termination.
The cell size in the $x$-direction was chosen to be $2a$ with $a=2.446$\AA\ which is the lattice constant of free standing Gr optimized in LDA.
We imposed the vacuum region in the $y$ and $z$ direction about 12\AA\ thick.
The cutoff energy of plane wave was chosen to be 400 eV and the $k$-points sampling was chosen to be $12\times1\times1$ with Monkhost-Pack mesh.
The geometry was optimized in the plane until maximum force becomes less than $10^{-3}$ eV/\AA\ since the out of plane displacements were relaxed into the flat structures.


First, we shall consider the stability of H termination in BCN nanoribbons.
The formation energy, $E_{\rm form}$, of BCN nanoribbon is defined as 
\begin{equation}
E_{\rm form} =E_{\rm BCN} - 2E_{\rm BN} 
- \frac{N_{\rm C}}{2}E_{\rm Gr} 
- \frac{N_{\rm H}}{2}E_{\rm H_2},
\end{equation}
where $E_{\rm BCN}$, $E_{\rm BN}$, $E_{\rm Gr}$ and $E_{\rm H_2}$ are the total energies of BCN nanoribbon, BN sheet, free standing Gr and isolated H$_2$ molecules, respectively.
Here, $N_{\rm C}$ and $N_{\rm H}$ are the number of C and H atoms in unit cell, respectively.
Corresponding spin distributions BCN nanoribbons within LSDA are presented in Fig.\ \ref{fg:BN-alt-Fig1} (c) by isosurface plots with isovalue $\pm 1.0\times 10^{-3}/{\rm bohr}^3$.
The same plot for GNR with H termination is shown in Fig.\ \ref{fg:BN-alt-Fig1} (c-v) as the reference.
The spin distribution at BH,NH, BH,N and B,NH structure edges are ferromagnetic while those at B,N and GNR edges are ferrimagnetic.
Calculated $E_{\rm form}$ for several structures with different magnetic configurations are summarized in Table 1.
We found that the magnetized BH,NH structure is stable.

In this comparison, however, effect of environment, i.e., temperature and pressure of H$_2$ gas is absent \cite{Wassmann2008prl}.
In order to clarify the effect, we shall consider the Gibbs free energy which is defined as $G =E_{\rm form}-N_{\rm H}\mu_{{\rm H}_2}/2$ with chemical potential of hydrogen gas $\mu_{{\rm H}_2}$ \cite{Wassmann2008prl}.
In Fig.\ \ref{fg:BN-alt-Fig1} (b), calculated Gibbs free energy measured from that of nonmagnetic B,N structure, $\Delta G$, is presented as a function of $\mu_{\rm H_2}$.
We used the stablest state for each structure for the calculation of $G$.
We found that magnetized BH,NH nanoribbon is stable in H$_2$ rich environment and nonmagnetic B,N nanoribbons is stable in H$_2$ poor environment.

\begin{table}[b!]
\centering
\caption{
Calculated results of formation energies at zero temperature $E_{\rm form}$ for several magnetic structures.
}
\begin{tabular}{c|cccc}
\hline \hline 
$E_{\rm form}$ (eV) & BH,NH & BH,N & B,NH & B,N \\
\hline 
LDA  &\quad $5.56$ \quad &\quad $9.20$ \quad &\quad $11.40$ \quad &\quad $13.29$ \quad \\
LSDA &\quad $5.54$ \quad &\quad $9.11$ \quad &\quad $10.79$ \quad &\quad $13.36$ \quad \\
\hline \hline 
\end{tabular}
\end{table}


Next, we shall consider the band structures of BCN nanoribbons.
The band structures of BCN nanoribbons within LDA are shown in Fig.\ \ref{fg:BN-alt-Fig1} (d).
In Fig.\ \ref{fg:BN-alt-Fig1} (d-v), the band structure of GNRs in the doubled unit cell within LDA are shown as the references.
We observed the partially flat bands at the Fermi level around the $\Gamma$ point in BCN nanoribbons with BH,NH and B,N structures, while the flat bands vanish if B and N atoms are uniformly substituted for outermost C atoms \cite{Kaneko2012submitted}.
The length of the flat bands of BCN nanoribbons in the wavevector space is same as that of GNR.
For GNR without H termination, the dangling bond states of C atoms are partially occupied, but flat bands are open when we do not passivate the dangling bonds \cite{Kawai2000prb}.
On the other hand, for BCN nanoribbon with B,N structure, the dangling bond states of N is 2.0 eV below, while those of B is 0.7 eV above the Fermi level, resulting in the appearance of the flat bands at the Fermi level.
For BCN nanoribbons with BH,N and B,NH structures, we could not observe the flat bands at the Fermi level but we found doubly degenerate bands consisting from $\pi$-orbital in $|k|<\pi/4a$ around the Fermi level.

Figure \ref{fg:BN-alt-Fig1} (e) shows the band structures of BCN nanoribbons within LSDA.
In Fig.\ \ref{fg:BN-alt-Fig1} (e-v), the band structure of GNRs in the doubled unit cell within LSDA are also presented as the references.
We found that BCN nanoribbons are semiconducting except the BH,N structure.
The band gaps of BCN nanoribbons are 0.155 eV for BH,NH structures and 0.189 eV for B,N structures, which are smaller than that of GNR 0.201 eV.
Although the band structures of BCN nanoribbons are similar to those of GNRs, the electronic properties and magnetic properties of these nanoribbons are quite different as discussed below.

In the remaining part of the letter, we shall analyze the flat band and edge states at the Fermi level observed in the BH,NH and B,N structures.
First, we shall use the tight binding model of $\pi$-electrons and the Hamiltonian is given by \cite{Yoshioka2003jpsj}
\begin{equation}
{\cal H} = \sum_{i} E_i c_{i}^\dagger c_{i} - 
\sum_{\langle i,j \rangle} t_{i,j} c_{i}^\dagger c_{j},
\end{equation}
where $E_i$ is an energy of $\pi$ electron at the site $i$, $c_{i}^\dag$ and $c_{i}$ are the creation and annihilation operators of electrons at the lattice site $i$, respectively, $\langle i,j\rangle$ stands for summation over the adjacent atoms and $t_{i,j}$ is the hopping integral of $\pi$ electrons from $j$th atom to $i$th atom.
$E_i$ are $E_{\rm B}$, $E_{\rm C}$ and $E_{\rm N}$, the site energies at the B, C and N atoms, respectively.
Following to the Yoshioka {\it et al}., we shall assume that the hopping integrals are constant regardless of the atoms, i.e., $t_{i,j}\equiv t$ and $E_{\rm B}>0$, $E_{\rm C}=0$, and $E_{\rm N}=-E_{\rm B}$. \cite{Yoshioka2003jpsj}

\begin{figure}[t!]
\centering
\includegraphics[width=7cm]{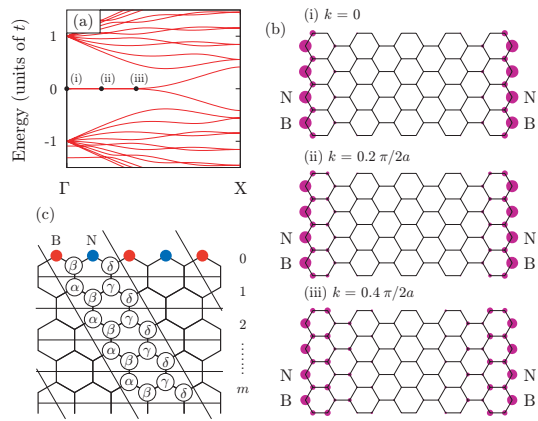}
\caption{
(color online)
(a) The band structure of BCN nanoribbon for $N=10$ and $E_{\rm B}/t=1$ within the tight binding model.
(b) The electronic charge distributions in BCN nanoribbon for the states indicated in Fig.\ \ref{fig:TB-1.0} (a).
(c) Schematics of semi-infinite Gr sheet with zigzag edge where the outermost C atoms are alternately substituted with B and N atoms.
}
\label{fig:TB-1.0}
\end{figure}

Calculated band structure of BCN nanoribbon for $N=10$ and $E_{\rm B}/t=1$ is shown in Fig.\ \ref{fig:TB-1.0} (a).
We reproduced partially flat bands at $E=0$ around $k=0$ which are quite similar to those within the first-principles calculations.
In Figs.\ \ref{fig:TB-1.0} (b), the charge distributions of BCN nanoribbon in the lowest conduction band at $k=0$ (i), $k=0.2\pi/2a$ (ii) and  $k=0.4\pi/2a$ (iii) are indicated by circles of which areas are proportional to the electronic charge.
At $k=0$, most of the electronic charges are accumulated at B and N atoms, and remaining electronic charges are distributed on sublattices belonging to the second outermost C atoms, which are resemble to those of the Klein edge \cite{Klein1994cpl}.
With increase of the wavevector $k$ as shown in Figs.\ (b-ii) and (b-iii), however, the electronic charges are distributed over both sublattices.
Therefore, the edge states at the BCN nanoribbons differ from conventional edge states.

The edge states can be analytically solved as follow:
We shall consider semi-infinite graphene with zigzag edge where outermost C atoms are alternately replaced with B and N as shown in Fig.\ \ref{fig:TB-1.0} (c).
Let $\lambda_{1}=-\eta+\eta^{-1}$ and $\lambda_{2}=-1/(\eta+\eta^{-1})$ with $\eta=e^{2ika}$.
Then, the amplitudes of $E=0$ state in $m$th low can be written as $\psi_{m,\alpha}= \lambda_1^m \psi_{\rm B}$, $\psi_{m,\beta}= -(E_{\rm B}/t) \lambda_2^{m+1} \psi_{\rm B}$, $\psi_{m,\gamma}= -\lambda_1^m \psi_{\rm B}$, and $\psi_{m,\delta}= -(E_{\rm B}/t) \lambda_2^{m+1} \psi_{\rm B}$, where $\psi_{\rm B}$ is the amplitude at B atom.
The edge state is defined as an evanescent state.
Since $|\lambda_1|<1$ for $|k|<\pi/3a$ and $|\lambda_2|<1$ in whole Brillouin zone, the flat bands exits $0\leq k < \pi/3a$.
As shown in Figs.\  \ref{fg:BN-alt-Fig1} (c-i) , \ref{fg:BN-alt-Fig1} (c-iv), and \ref{fig:TB-1.0} (a), this result is consistent with numerical results.

Next, we shall discuss the change in the magnetic structures in BCN nanoribbons using the Hubbard model which is constructed by adding the on-site Coulomb interaction term to the tight binding model.
The Hamiltonian of the system is given by
\begin{equation}
{\cal H} = \sum_{i,\sigma} E_i c_{i,\sigma}^\dagger c_{i,\sigma} 
- t \sum_{\langle i,j \rangle,\sigma} c_{i,\sigma}^\dagger c_{j,\sigma} 
+ U \sum_{i} n_{i,\uparrow}n_{i,\downarrow}, 
\label{eq:Hubbard}
\end{equation}
where $U$ represents strength of on-site Coulomb interaction.
In this study, we shall assume that the on-site Coulomb interaction is independent on the chemical species of the site $i$.
We adopted the mean-field approximation of the Hamiltonian Eq.\ (\ref{eq:Hubbard}).

\begin{figure}[t!]
\centering
\includegraphics[width=7cm]{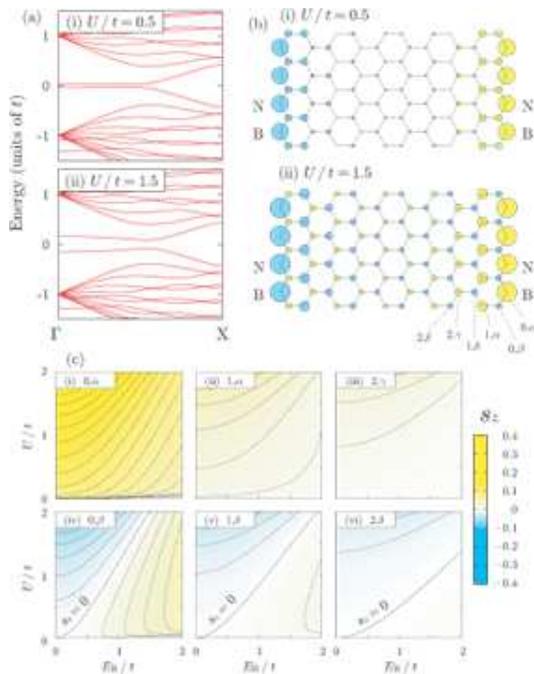}
\caption{
(color online)
(a) The band structure of BCN nanoribbons with $N=10$ using the Hubbard model for $U/t = 0.5$ (i) and for 1.5 (ii).
(b) The spin distribution for $U/t = 0.5$ (i) and for 1.5 (ii).
In these figures, up and down spin density distributions are represented by the yellow (positive) and light blue (negative) circles whose areas are proportional to the magnitude of spin density at the site. 
(c) Color plots of $E_{\rm B}$ and $U$ dependence of $s_z$ at several sites shown in the bottom of (b-ii).
The site indexes are consistent with those defined in Fig.\ \ref {fig:TB-1.0} (c).
In each panel, the solid line represents $s_z=0$.
}
\label{fig:Hubbard-Band-Spin}
\end{figure}

The band structures for $E_{\rm B}/t=1$ are presented in Fig.\ \ref{fig:Hubbard-Band-Spin} (a) for $U/t=0.5$ (i) and for $1.5$ (ii).
We obtained spin split energy bands which agree well those within LSDA calculations shown in Figs.\ \ref{fg:BN-alt-Fig1} (e-i) and (e-vi).
In Fig.\ \ref{fig:Hubbard-Band-Spin} (b) the distributions of spin densities for $E_{\rm B}/t=1$ are presented by the circles whose areas are proportional to the magnitude of spin density for $U/t=0.5$ (i) and for $1.5$ (ii).
We found that ferrimagnetic order is realized for $U/t=0.5$, but ferromagnetic order is realized for $U/t=1.5$.
In Fig.\ \ref{fig:Hubbard-Band-Spin} (c) $E_{\rm B}$ and $U$ dependences of spin at several sites are presented in color plot.
In this figure, $s_z=0$ is indicated by the white solid line and the contours are indicated by the dashed lines.
It should be noted that spin density at $m,\alpha$ and at $m,\beta$ site equals to those at $m,\gamma$ at $m,\delta$ site, respectively.
From these figures, we found that the magnetic order depends on $U/E_{\rm B}$, i.e., ferromagnetic order is realized for $U/E_{\rm B}\ll1$ and ferrimagnetic order is realized for $U/E_{\rm B}\gg1$.

Above results within LSDA and the Hubbard model suggest that $E_{\rm B}$ increases and $E_{\rm N}$ decrease when outermost atoms are terminated by H atoms.
To understand this behavior, we consider the shift of average potential at core, $\Delta\bar{V}$, by hydrogenation, defined as $\Delta\bar{V} = (\bar{V}^{(\rm BH,NH)}- E_F^{(\rm BH,NH)}) - (\bar{V}^{(\rm B,N)}- E_F^{(\rm B,N)})$, where $\bar{V}$'s are the averaged potentials at the atomic cores and $E_F$'s are the Fermi energies.
We obtained $\Delta\bar{V}$ within LDA 2.05 eV for B and $-1.99$ eV for N.
As the reference we also obtained $-0.09$ eV for the outermost C atoms in GNR.
We found that the core levels shift upward for B atoms and downward for N atoms by passivation with H atom, resulting in the considerable increase in $E_{\rm B}$, since the hopping integral $t$ in Gr is 2.7 eV.
These large shift of the energies $\pi$-electron in B and N atoms causes change of magnetic structure from ferrimagnetic to ferromagnetic order in the edge states.

In summary, we investigated the electronic and magnetic properties of boron-carbon-nitride (BCN) nanoribbons with zigzag edges where the outermost C atoms on the edges are alternately replaced with B and N atoms using the first principles calculations.
We showed that BCN nanoribbons in both H$_2$ rich and poor environment have the partially flat bands at the Fermi level around the $\Gamma$ point.
The band structures of BCN nanoribbons are quite similar to those in graphene nanoribbons in doubled unit cell but the distribution of charge and spin densities are different from those in graphene nanoribbons.
We analyzed the edge states corresponding to the flat bands using the tight binding model and the Hubbard model in order to clarify the origin of difference from those in graphene nanoribbons.
We showed that the origin of such different charge and spin density distributions is the different site energies of B and N atoms compared with C atoms.

{\it Acknowledgments} - 
The authors acknowledge Y.\ Shimoi, H. Arai, H. Tsukahara, K.\ Wakabayashi, S.\ Dutta and M.\ B\"urkle for valuable discussions.
This research was supported by the International Joint Work Program of Daeduck Innopolis under the Ministry of Knowledge Economy (MKE) of the Korean Government.

\end{document}